\documentclass[oldversion,letter]{aa}
\usepackage{natbib}
\usepackage{txfonts}
\usepackage{graphicx}
\bibpunct{(}{)}{;}{a}{}{,} 

\def\oversim#1#2{\lower0.5pt\vbox{\baselineskip0pt \lineskip-0.5pt
     \ialign{$\mathsurround0pt #1\hfil##\hfil$\crcr#2\crcr\sim\crcr}}}
\def\gsim{\mathrel{\mathpalette\oversim>}}    
\def\lsim{\mathrel{\mathpalette\oversim<}}    
\def\nt#1{\vtop{\tiny\hsize=\columnwidth\leavevmode#1\hspace*{\fill}}}

\def\er{.\phantom{0}}
\def\eer{.\phantom{0 }}
\def\cn#1{\multicolumn{1}{c}{#1}}

\begin{document}

\title{The onset of photoionization in Sakurai's Object (V4334 Sgr)\thanks{Based
on observations collected at the European Southern Observatory, Chile (programmes
71.D-0396, 75.D-0471, 77.D-0394).}}

\author{P.A.M.~van Hoof \inst{1}
  \and M.~Hajduk \inst{2}
  \and Albert A.~Zijlstra \inst{3,4} 
  \and F. Herwig \inst{5}
  \and A.~Evans \inst{5}
  \and G.C Van de Steene \inst{1}
  \and S. Kimeswenger \inst{6}
  \and F.~Kerber \inst{7}
  \and S.P.S.~Eyres \inst{8}
      }

\institute{
  Royal Observatory of Belgium, Ringlaan 3, 1180 Brussels, Belgium
  \and
  Centrum Astronomii UMK, ul.Gagarina 11, PL-87-100 Torun, Poland 
  \and
  University of Manchester, School of Physics \&\ Astronomy, P.O. Box 88,
  Manchester M60 1QD, UK
  \and
  South African Astronomical Observatory, P.O. Box 7935, Observatory, South Africa
  \and  
  Astrophysics Group, School of Physical and Geographical Sciences, Keele
  University, Staffordshire ST5 5BG, UK
  \and
  Institute of Astro and Particle Physics, University Innsbruck, 
  Technikerstra{\ss}e 25, 6020 Innsbruck, Austria
  \and
  European Southern Observatory,
  Karl-Schwarzschild-Stra{\ss}e 2, D-85748 Garching, Germany
  \and
  Centre for Astrophysics, University of Central Lancashire, Preston PR1 2HE, UK
  }

\date{Accepted Received; in original form }

\abstract{We investigate the reheating of the very late thermal pulse (VLTP)
  object V4334 Sgr (Sakurai's Object) using radio observations from the Very
  Large Array, and optical spectra obtained with the Very Large Telescope. We
  find a sudden rise of the radio flux at 5 and 8~GHz --- from $\leq 90~\mu$Jy
  and $80~\pm 30~\mu$Jy in February 2005 to $320~\mu$Jy and $280~\mu$Jy in
  June 2006. Optical line emission is also evolving, but the emission lines
  are fading. The optical line emission and early radio flux are attributed to
  a fast shock (and not photoionization as was reported earlier) which
  occurred around 1998. The fading is due to post-shock cooling and
  recombination. The recent rapid increase in radio flux is evidence for the
  onset of photoionization of carbon starting around 2005. The current results
  indicate an increase in the stellar temperature to 12~kK in 2006. The mass
  ejected in the VLTP eruption is $M_{\rm ej} \geq 10^{-4}~\rm M_\odot$, but
  could be as high as $10^{-2}~\rm M_\odot$, depending mainly on the distance
  and the clumping factor of the outflow. We derive a distance between 1.8 and
  5~kpc. A high mass loss could expose the helium layer and yield abundances
  compatible with those of [WC] and PG1159 stars.}


\maketitle

\section{Introduction}

Helium shell flashes dominate many aspects of the evolution of AGB (Asymptotic
Giant Branch) and post-AGB stars. Helium flashes (or thermal pulses) are
common on the AGB but the effects on the surface properties are mitigated by
the thick hydrogen envelope. Very late thermal pulses (VLTPs), which can occur
after hydrogen burning has ceased, lead to large and rapid changes in the star
and are the best way to constrain the physics of these thermonuclear
eruptions. The VLTP in Sakurai's Object (V4334 Sgr) has provided the first
opportunity to observe such an event with modern instrumentation, and allows
us to study poorly understood aspects, such as the post-VLTP mass loss and the
mixing length theory \citep{Hajduk2005}.

Very few VLTP events have been observed: only V4334 Sgr \citep{Duerbeck1997}
and V605 Aql \citep{Clayton2006, Lechner2004} were discovered during their
high-luminosity phase (respectively in 1996 and in 1918). CK Vul (in 1670) is
suspected to represent a third case \citep{Hajduk2007}. FG Sge also shows
evidence for current post-VLTP evolution but its status is uncertain
\citep{Jeffery2006}. Another five central stars of planetary nebulae (PNe)
show evidence for historical VLTP eruptions, based on the presence of
hydrogen-poor gas near the central star \citep{Zijlstra2002}.

V4334 Sgr showed a high luminosity phase with a cool stellar atmosphere
($T_{\rm eff} \sim 6000~\rm K$) within a few years after the VLTP, during
which the star became hydrogen poor and carbon rich. The speed of evolution
from the VLTP to maximum brightness was surprising as pre-Sakurai models
predicted timescales of\ \,$>$~100~ yr. The much faster evolution has been
explained as caused by suppressed convection under explosive conditions
\citep{Lawlor2003, Hajduk2005}. However, \citet{MillerBertolami2006} find that
the numerical time step is important, and that smaller time steps are
sufficient to predict much faster evolution. The problem of convection is
studied further by \citet{Herwig2006}. The reheating timescale after maximum
brightness is an important constraint to test the models. This evolution can
be traced best by observing the ionization of the surrounding nebula.
\citet{Eyres1999}, \citet{Tyne2000}, and \citet{Kerber2002} detected emission
from atomic and ionized species around V4334 Sgr which is evidence for
ionization. The existence of ionization was confirmed by \citet{Hajduk2005}
based on radio observations. In this Letter, we present new radio and optical
observations, showing that this early ionization was due to a fast shock.
Evidence for the onset of photoionization of carbon is also found, indicating
that this started around 2005. We present photoionization models of the ejecta
and compare the results to various VLTP evolutionary tracks.

\section{Observations}

\paragraph{Radio emission}

We have been monitoring the radio flux of V4334 Sgr to test the temperature
evolution of the star. Observations were carried out with the VLA array
between 2004 and 2006, at 5~GHz and 8~GHz. We observed in the hybrid BnC
(2004) and AnB (2005, 2006) arrays. This reduced confusion with emission from
the old, extended PN. Integration times were 3-5 hours, with 1331+305 as the
flux calibrator and 1733$-$130 as the phase calibrator. The assumed flux
density of 1331+305 was 4.66 Jy at 5 GHz. Flux calibration was interpolated to
V4334 Sgr. Observational parameters are summarized in Table 1. The resultant
visibilities were Fourier transformed to the image plane and the dirty beam
deconvolved. Convolution with a gaussian beam fitted to the dirty beam size
gave images from which fluxes were measured. Natural weighting was used for
the imaging. The FWHM of the gaussian beams are listed in Table 1.

\paragraph{Optical spectroscopy}

Line emission from the central ejecta in the optical wavelength regime was
first detected in 2001 by \citet{Kerber2002} using the FORS1 instrument on the
VLT. We continued to monitor the object with FORS1/2 in subsequent years.
Details are given in Table~\ref{lines}. We averaged observations taken within
the same semester. The spectra were manually reduced in the standard way using
{\sc iraf}. Sky lines were removed using the part of the long slit spectrum
outside the old PN. The old PN also contributes line emission: this is
strongly dependent on position on the slit, and the old PN lines were removed
by interpolation of the area around the central source. The response
correction was done using various white dwarf flux standards. The line fluxes
derived from the VLT spectra are listed in Table~\ref{lines}. The data were
not corrected for reddening. The 2001 values are taken from
\citet{Kerber2002}.

\section{Discussion}

\subsection{Extinction distance}

The FORS long slit spectra were used to extract the spectrum of the old PN,
taking care not to include any field stars or the recent ejecta. The
H$\alpha$/H$\beta$ ratio gives the interstellar extinction. Assuming a case B
ratio of 2.85, we derive $E({\rm B-V})=0.86$, somewhat higher than found by
\citet{Pollacco1999}. The extinction-distance diagram for stars within a few
arcminutes from V4334 Sgr is derived by \citet{Kimeswenger1998}: it shows a
linear rise with distance up to 1.8~kpc, followed by a constant extinction of
$E({\rm B-V})=0.9\pm0.09$ between 1.8 and 5~kpc.

This suggests that V4334 Sgr is located beyond 1.8~kpc. The previously used
value of 1.9~kpc \citep[e.g.][]{Evans2006} can be considered a lower limit.
\citet{Duerbeck2000} suggest $d \sim 3$--$5.4~$kpc based on the stellar
luminosity. The line of sight reaches the scale height of the old disk at $d
\approx 4~\rm kpc$ and this would provide a plausible distance.

\subsection{Shock ionization}

\begin{figure}
\includegraphics[width=62mm]{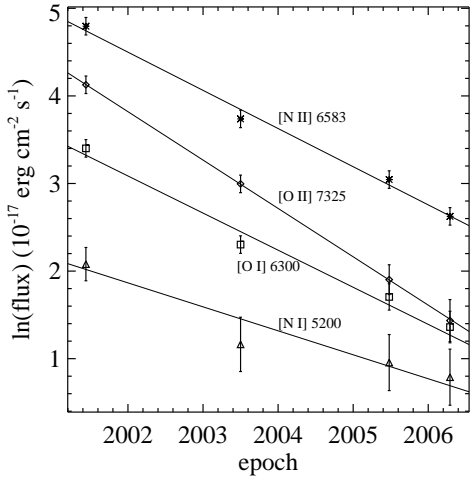}
\caption{\label{lineplot}Evolution of emission line strength. The lines
show the least-squares fit to the data.}
\end{figure}

The line fluxes in Table~\ref{lines} show a strong exponential decline with
time (see also Fig. \ref{lineplot}). The level of excitation is also
decreasing, as the high excitation [O\,{\sc ii}] lines are decreasing faster
than lines of lower excitation. The e-folding times are 1.81$\pm$0.13~yr for
[O\,{\sc ii}] (the highest excitation line), 2.29$\pm$0.14~yr for [N\,{\sc
  ii}], 2.4$\pm$0.2~yr for [O\,{\sc i}], and 3.7$\pm$0.9~yr for [N\,{\sc i}].
This is clearly inconsistent with photoionization by an increasingly hot
central star \citep{Kerber2002, Hajduk2005}. The decrease in intensity as well
as the level of excitation is consistent with a single shock that occurred
somewhere before 2001 and then started cooling and recombining
\citep[cf.][]{Kafatos1973}. The [N\,{\sc ii}] line ratio in 2001 indicates an
electron temperature $ T_{\rm e} = 3200 - 5500$~K, depending on the electron
density (Table~\ref{cloudy}). The [C\,{\sc i}] ratio in 2003 indicates $T_{\rm
  e} = 2300 - 4300$~K. These low values indicate that cooling had already
commenced by 2001.

\begin{table}
  \caption[]{\label{vla}Observational parameters and results for the VLA observations.
    PA gives the position angle of the beam.
  }
\begin{tabular}{llllllllll}
\hline
 Date & Freq. & FWHM Beam, PA & $\sigma$    & flux \\ 
      & [GHz] &               & [$\mu$Jy/b] & [$\mu$Jy] \\
\hline
5 Feb 2004 &   8 & 1.30\arcsec$\times$1.14\arcsec, +48$^\circ$ & 10 & $100\pm30$ \\
4 Feb 2005 &   5 & 1.37\arcsec$\times$0.96\arcsec, +68$^\circ$ & 16 & $<90$ \\
6 Feb 2005 &   8 & 0.78\arcsec$\times$0.54\arcsec, +76$^\circ$ & 14 & $70\pm40$ \\
11 Jun 2006 &  8 & 0.73\arcsec$\times$0.65\arcsec, +46$^\circ$ & 17 & $280 \pm 50$ \\
12 Jun 2006 &  5 & 1.43\arcsec$\times$1.05\arcsec,  +2$^\circ$ & 19 &  $320 \pm 60$ \\
\hline
\end{tabular}
\vspace*{-2pt}
\end{table}

\begin{table}
  \caption[]{\label{lines}Line fluxes in units of $10^{-17}~\rm
    erg\,cm^{-2}\,s^{-1}$. The 2001 fluxes are taken from \citet{Kerber2002}. 
    {\it OR} means outside the observable range.
    Where appropriate 3-$\sigma$ upper limits are given.}
\begin{tabular}{lr@{\hspace{2mm}}rrr}
\hline
Line           &  \llap{20}01.44$^a$ & 2003.50$^b$ & 2005.48$^c$ & 2006.29$^d$ \\
\hline
{}[N\,{\sc i}] 5198, 5200  &   8\er     &  3.2  $\pm$ 0.5 &  2.6  $\pm$ 0.5 &  2.2 $\pm$ 0.7 \\
{}[N\,{\sc ii}] 5755       &   2\er     &  2.1  $\pm$ 0.7 &   \cn{$<$ 1.5}  &   \cn{$<$ 2.2} \\
{}[O\,{\sc i}] 6300        &  30\er     & 10\eer$\pm$ 1.5 &  5.5  $\pm$ 0.8 &  3.9 $\pm$ 0.8 \\
{}[O\,{\sc i}] 6364        &   8\er     &  4\eer$\pm$ 1.2 &  2.5  $\pm$ 0.6 &  2.1 $\pm$ 0.7 \\
{}[N\,{\sc ii}] 6548       &  40\er     & 15\eer$\pm$ 1.9 &  7.6  $\pm$ 0.9 &  3.7 $\pm$ 0.6 \\
{}[N\,{\sc ii}] 6583       & 121\er     & 42\eer$\pm$ 4.2 & 21\eer$\pm$ 2.2 & 13.8 $\pm$ 1.5 \\
{}[S\,{\sc ii}] 6716       &   2.5      &  \cn{$<$ 1.9}    &   \cn{$<$ 1.6}  &   \cn{$<$ 1.3} \\
{}[S\,{\sc ii}] 6731       &   2\er     &  \cn{$<$ 1.9}    &   \cn{$<$ 1.6}  &   \cn{$<$ 1.3} \\
{}[O\,{\sc ii}] 7319, 7330 &  62\er     & 20\eer$\pm$ 2.1 &  6.7 $\pm$ 1.6  &  4.2 $\pm$ 1.0 \\
{}[C\,{\sc i}] 8727        &\cn{\it OR} & 13\eer$\pm$ 1.5 &   \cn{$<$ 3.9}  &  4.7 $\pm$ 0.8 \\
{}[C\,{\sc i}] 9824, 9850  &\cn{\it OR} &150\eer$\pm$ 15. &   \cn{\it OR}   &   \cn{\it OR}  \\
\hline \\
\vspace*{-16pt}
\end{tabular}
\nt{$^a$~300V+GG375, 4500 -- 8000~\AA. $^b$~150I, 4900 -- 10075~\AA;
150I+OG590, 6200 -- 10075~\AA; 1200R+GG435, 5875 -- 7365~\AA.
$^c$~300V+GG435, 4800 -- 8900~\AA. $^d$~600V+GG435, 4520 -- 6850~\AA;
600I+OG590, 6500 -- 9200~\AA.}
\end{table}

The origin of the shock is not clear. No emission lines were seen during 1997.
The earliest evidence for shock ionization is based on the He\,{\sc i}
recombination line, and dates from 1998 \citep{Eyres1999, Tyne2000}.
High-velocity CO was seen starting late 1999 or early 2000 \citep{Eyres2004},
with an outflow velocity around 290 km\,s$^{-1}$. Optical shock emission with
velocities up to $-$350 / $+$200~km\,s$^{-1}$ was first seen in 2001
\citep{Kerber2002}. The shock occurred close to the time of the onset of dust
formation \citep{Duerbeck2000}. Neutral molecule chemistry is slow in a
hydrogen-deficient environment, while ion-molecule chemistry is much faster
\citep{Woods2002}. Hence UV radiation from the shock may have triggered dust
formation elsewhere in the nebula through ionization of key molecular species.
If the ejecta are bipolar, the dust formation could be taking place in the
dense equatorial region (possibly an equatorial disk), while the shocks are
internal to a fast polar outflow \citep[]{Evans2006}.

\subsection{Photoionization}

The radio flux was constant or decreased slowly between 2004 and 2005. This is
consistent with the model of an early fast shock: the radio data trace the
post-shock cooling and recombination of the shock and not a photoionized
region. However, the 2006 observations show a strong increase in radio flux,
by a factor of four within 16 months. The only optical line which appears to
have strengthened is the [C\,{\sc i}] 8727 line, but this is uncertain due to
the low S/N of the 2005 data. The increase in radio flux in 2006 is therefore
attributed to newly photoionized carbon which has the lowest ionization
potential of all the major constituents in the gas. This process would not
produce any noticeable changes in the optical or mid-IR spectrum.

We ran Cloudy models to investigate the origin of the radio flux using version
c07.04.01 of the code, last described by \citet{Ferland1998}. The basic models
are similar to those described in \citet{Hajduk2005}, with the following
changes. (1) The angular diameter of the dust shell is discussed in
\citet{Tyne2002} and \citet{Evans2006}. Extrapolating their data to 2006 leads
to an angular diameter of $\sim$ 100~mas. However, with such a small diameter
the plasma would become optically thick at 5~GHz before the necessary emission
measure could be reached. Based on the FORS2 1200R spectral image taken in
2003 (seeing 0.54\arcsec) we measured a deconvolved (gaussian) FWHM of
0.27\arcsec\ for the [N\,{\sc ii}] emitting region. This indicates that the
true diameter must be 0.3 -- 0.5 arcsec \citep{vanHoof2000}. We modeled both
limits of this range. (2) Abundances from \citet[October 1996
values]{Asplund1999} were used, except for carbon where we used C/He = 0.1 by
number (the value \citet{Asplund1999} used for their model atmospheres). (3)
For the stellar spectrum we used the models of \citet{Castelli2004}. (4) We
assumed a clumpy medium with a filling factor $f$ of 0.1 or 0.01, and a
$1/r^2$ density distribution. (5) The total dust mass was the same in all
models with the same distance (see Sect.~\ref{shellmass}). We list four models
in Table~\ref{cloudy}. The implied gas-to-dust mass ratio is $\sim$
580\,$\sqrt{f}$ for the $d=1.9$~kpc models and $\sim$ 870\,$\sqrt{f}$ for the
$d=4$~kpc models. The ionization of carbon begins around $T_{\rm eff}= 10~\rm
kK$. The radio flux (without the shock emission) as a function of $T_{\rm
  eff}$ is listed in Table~\ref{temp}. The observed radio flux is consistent
with an increase in $T_{\rm eff}$ from $\lsim$ 11~kK in 2005 to $\sim$ 12~kK
in 2006.

\subsection{Stellar evolution models}

Three evolutionary tracks have been published which can fit the rapid post
He-flash evolution. \citet{Herwig2001} and \citet{Hajduk2005} proposed
suppressed convection after the flash. \citet{Lawlor2003} showed that with
this assumption, the evolution could be closely matched; they first predicted
the 'double loop' evolution. \citet{MillerBertolami2006} find that the initial
fast changes can also be found in models with standard convection, by choosing
a very small time step in the simulations. This raises the issue of the
accuracy of the previous models. In all cases the initial fast evolution is
caused by a hydrogen ingestion flash at the top of the helium layer.

\begin{table}
\caption[]{\label{cloudy}Cloudy model parameters. The main constituents of the gas are helium (He/H = 250 by number) and carbon (C/H = 25 by number).
The electron density in the ionized region
is approximately 25\,$n_{\rm H}$. $ M_{\rm d}$ and $ M_{\rm ej}$ are the dust mass and the total ejecta mass.}
\begin{tabular}{llllllllll}
\hline
 & {model 1} &  {model 2}  & {model 3} &  {model 4} \\
\hline 
 $d$ [kpc]       & 1.9 & 1.9 & 4.0 & 4.0 \\   
 $L$ [L$_\odot$] & 2770 & 2770 & 12280 & 12280 \\
 $\Theta$ [arcsec] & 0.3 & 0.5 & 0.3 & 0.5 \\
 $\log(r_{\rm in} [\rm cm])$ & 14.931 & 15.153 & 15.254 & 15.476 \\
 $\log(r_{\rm out} [\rm cm])$ & 15.630 & 15.852 & 15.953 & 16.175 \\
 $\log(n_{\rm H} [\rm cm^{-3}]/\sqrt{f})$ & 4.45 & 3.85 & 4.3 & 3.7 \\
 $M_{\rm d}$ [$10^{-6}$ M$_\odot$] & 1.95 & 1.95 & 8.65 & 8.65 \\
 $M_{\rm ej}/\sqrt{f}$ [$10^{-3}$ M$_\odot$] & 1.05 & 1.22 & 6.92 & 8.05 \\
\hline
\end{tabular}
\end{table}

\begin{table}
\caption[]{\label{temp}Predicted radio flux (not including the shock emission) at 8.435~GHz in $\mu$Jy
(valid both for $f$ = 0.1 and 0.01).}
\begin{tabular}{llllllllll}
\hline
$T_{\rm eff}$ & {model 1} & {model 2} & {model 3} & {model 4} \\
\hline 
10000    &    4 &    4 &    4 &    3 \\
11000    &   24 &   24 &   24 &   24 \\
12000    &  340 &  330 &  350 &  340 \\
13000    &  830 &  790 &  850 &  820 \\
\hline
\end{tabular}
\end{table}

\begin{figure}
\includegraphics[width=77mm]{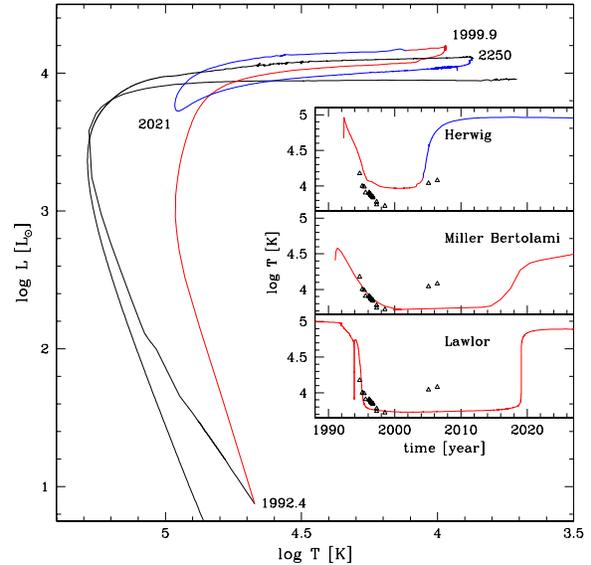}
\caption{\label{evo}Comparison of predicted and observed 
evolutionary timescales.}
\end{figure}

Fig. \ref{evo} shows the double-loop model of \citet{Herwig2001} and
\citet{Hajduk2005}, with the top inset showing the temperature evolution
during the first return. The middle inset shows the model of
\citet{MillerBertolami2006} and the lower inset shows a representative model
of \citet{Lawlor2003} (not necessarily their best-fitting one). The
observational data points come from \citet{Duerbeck1997}, \citet{Asplund1999},
and \citet{Pavlenko2002}. The last two points are from this paper. For
broadband photometric data (including pre-discovery photometry), we have
estimated temperatures from bolometric corrections. These corrections are
based on normal giant photospheres and are therefore uncertain. The low
temperature between 1995 and 2000 is caused by an expanding pseudo-photosphere
\citep{Duerbeck2000}. The underlying star likely remained hotter. Such an
expanding photosphere is difficult to model with a stellar evolution code and
predictions for $T_{\rm eff}$ in this phase are therefore uncertain. However,
this does affect the overall reliability of the track.

The three models differ in the predictions for the reheating timescale, but
neither fully fits the current observations. Herwig's model reheats faster,
and the other two models considerably slower than the observations indicate.
The reheating timescale in Herwig's model can be adjusted by tuning the core
mass or the mixing efficiency in the He-flash convection zone
\citep{Herwig2001}. There is a degeneracy between these two parameters and the
effects of a smaller core mass and reduced mixing efficiency are to some
degree interchangeable. The mass loss during the cool phase may also be an
important missing ingredient in these models \citep{Iben1995}.

\subsection{Ejecta mass}
\label{shellmass}

The results from the Cloudy modeling can be summarized in the following
formula for the ejected mass:
\begin{equation}
  M_{\rm ej} = 6.0\ 10^{-4} \left(\frac{f}{0.01}\right)^{0.5}
  \left(\frac{d}{4\ \rm kpc}\right)^{2.5}
  \left(\frac{\Theta}{0.5\arcsec}\right)^{0.3} \frac{m(\rm He)}{2.5} \ {\rm
    M_\odot}, 
\end{equation}
where $\Theta$ is the diameter of the nebula and $m(\rm He)$ is the He/C mass
ratio. The dominant uncertainty in the total ejecta mass comes from the
distance, the volume filling factor, and the He/C mass ratio. In the Herwig
models, the He/C ratio quickly evolves to a value of 2.5 after the star has
reached the red end of the first loop. We take this value to determine the
minimum ejecta mass, required by the photoionization models. Hence a very
conservative lower limit to the ejecta mass is $6\ 10^{-4}~\rm M_\odot$ at
$d=4$~kpc and $10^{-4}~\rm M_\odot$ at $d=1.9$~kpc assuming extreme clumping
($ f = 0.01$). For a uniform outflow the lower limits would be a factor of 10
higher. The corresponding mass-loss rate over a period of roughly a decade is
very high. It could be comparable to, or even exceed, the mass-loss rate
during the superwind phase. A minimum ejecta mass of $ 5\ 10^{-3}~\rm M_\odot$
is sufficient to expose the intershell region \citep[e.g.\ Fig. 3
in][]{Iben1995} where the C/He ratio reaches its peak value of 2.66 by mass.
This would leave a central star with abundances compatible with those of [WC]
and PG1159 stars. The linear momentum in the ejecta far exceeds that in the
stellar radiation (even when considering multiple scattering), so that the
ejecta must be energy driven. This suggests that the ejecta may have been lost
instantaneously.

The dust mass determined from the continuum emission has been steadily
increasing, starting in 1998 or early 1999 \citep{Tyne2002}. In June 2003, the
total dust mass indicated by the sub-mm continuum was $1.9 \times 10^{-6}~\rm
M_\odot$ at $d=4$~kpc. Between January and June 2003 the dust mass increased
by a factor of 1.7. The increase was used to derive a mass-loss rate of
$1.4\times 10^{-4}~\rm M_\odot\, yr^{-1}$ at $d=4$~kpc by 2003
\citep{Evans2004}. However, the concept of converting the dust growth rate to
a mass-loss rate assuming a constant and high dust-to-gas ratio (1/75 by mass)
can be questioned. The low observed dust temperature, $T_{\rm d} \approx
400~\rm K$ in 2003 seems difficult to reconcile with the premise that half the
dust was newly formed and therefore located close to the dust condensation
radius. Instead, the increase in dust mass can represent a continuous dust
formation or growth in a circumstellar disk \citep[cf.][]{Hajduk2007}. Based
on the flux in the Spitzer spectrum \citep{Evans2006}, we find a current dust
mass $ M_d \approx 8.7 \times 10^{-6}~\rm M_\odot$ at $ d=4$~kpc, using
opacities for graphite. The dust mass has increased fivefold between June 2003
and April 2005 (when the Spitzer data were taken). The average dust growth
rate during this time was $\dot M_d \approx 3.7 \times 10^{-6}~\rm
M_\odot\,yr^{-1}$, a factor of 2 higher than in 2003.

\section{Conclusions}

The photoionization of carbon in the ejecta of V4334 Sgr has started around
2005. This is evident from a 4-fold increase in the 8~GHz radio flux observed
in June 2006. Simultaneously the optical spectrum shows a continuing
exponential decline, both in excitation and ionization. This indicates that
the emission lines are formed in a shock which occurred around 1998 and is
currently cooling and recombining. Based on the Balmer decrement in the
spectrum of the old PN, we could determine that the distance to V4334 Sgr is
in the range 1.8 -- 5~kpc, with a preferred value $ \sim$ 4~kpc. New Cloudy
models of the photoionized region indicate that the total shell mass is at
least $10^{-4}$~M$_\odot$, but could be as high as $10^{-2}$~M$_\odot$
(assuming $ f=1$ and $ d=5$~kpc), depending mainly on the assumed distance and
clumping factor in the outflow. The models indicate that the star was at a
temperature of 12~kK in 2006 and is currently evolving with a speed $\gsim$
1~kK\,yr$^{-1}$. This confirms that V4334 Sgr not only displays a fast return
to minimum temperature, but also a fast reheating. A comparison of the
temperature evolution of the central star with observations shows that
Herwig's model reheats faster \citep{Herwig2001, Hajduk2005}, and the other
two models considerably slower than the observations indicate
\citep{MillerBertolami2006, Lawlor2003}. The dust-to-gas mass ratio may be
lower than hitherto assumed, depending mainly on the clumping factor of the
gas. Our analysis indicates that dust formation and/or growth is slow and
still ongoing.

\section*{Acknowledgments}

PvH acknowledges support from the Belgian Science Policy Office (grant
MO/33/017). MH acknowledges support from the Polish research grant No. N203
024 31/3879. We wish to thank R. Blomme and D. Jevremovi\'c for stimulating
discussions.

\bibliographystyle{aa} 
\bibliography{7932} 

\end{document}